\documentstyle[preprint,aps]{revtex}
\begin{document}
\date{\today}
\twocolumn
\tighten 
\title{The effects of compressible and incompressible states on the
       FIR-absorption of quantum wires and dots in a magnetic field.}
\author{Vidar Gudmundsson$^1$, Arne Brataas$^2$, Christoph Steinebach$^3$,\\ 
        A. G. Mal'shukov$^4$, K. A. Chao$^2$, and Detlef Heitmann$^3$}

\address{$^1$ Science Institute, University of Iceland, Dunhaga 3,
         \\ IS-107 Reykjavik, Iceland.\\ 
         $^2$ Department of Physics,
         Norwegian University of Science and Technology, \\ N-7034
         Trondheim, Norway.\\ 
         $^3$ Institut f{\"u}r Angewandte Physik, Jungiusstra{\ss}e 11,
         D-20355 Hamburg, \\ Federal Republic of Germany.\\
         $^4$ Institute of Spectroscopy, Russian
         Academy of Sciences, 142092 Troitsk, \\ Moscow Region, Russia.}
\maketitle
\begin{abstract}
      We investigate the effects of compressible and incompressible 
      states on the FIR-absorption of quantum wires and dots in a
      homogeneous perpendicular magnetic field.
      The electron-electron interaction is treated in the Hartree
      approximation at a finite low temperature. 
      The calculated dispersion of the collective excitations reproduces
      several experimental results.

\end{abstract}

\section{Introduction}
According to Kohn's theorem 
\cite{Kohn61:1242,Maksym90:108,Bakshi90:7416,Pfannkuche93:6} a spatially
homogeneous electric field can only excite the rigid center of mass motion of
all the electrons confined parabolically in a quantum dot or wire. 
In far-infrared (FIR) spectroscopy this condition is usually satisfied due to 
the small size of the individually isolated dots and wires
compared to the wavelength of the radiation. 
In order to detect some of the internal structure or relative motion of the
two-dimensional electron gas (2DEG) through FIR spectroscopy
the confining potential has to deviate from the parabolic form.
The deviation is found to influence the FIR response of the 2DEG in two 
different ways, besides a trivial blue or red shift:
First, the center of mass mode (the magnetoplasmon)
interacts with higher order harmonics of the cyclotron resonance forming
complicated anticrossing patterns \cite{Demel90:788a,Gudmundsson95:17744}.
Second, the formation of compressible and incompressible stripes 
\cite{Beenakker90:216,Chklovskii92:4026,Lieb95:10646}
in the electron density modulates slightly the dispersion of the lower
lying branch in quantum dots \cite{Bollweg96:2774}.
In addition, effects of the stripe formation in quantum 
dots have been measured both in tunneling \cite{Vaart94:320} and 
transport experiments \cite{Stopa96:2145}.

In this report we elucidate further the connection between the
modulation of the lower magnetoplasmon branch in quantum dots
and the formation of compressible and incompressible stripes
in the electron density. We explain the apparent weakness of this 
phenomenon in the upper main magnetoplasmon branch in quantum dots
and the main branch in quantum wires. In both systems we find
similar modulation in the dispersion of weak higher order 
collective oscillations.

In order to have consistent results for the absorption we use
a corresponding self-consistent method for the ground state properties
and the excited states of the 2DEG.

\section{Model}
The ground state properties of a single quantum dot or a wire
are calculated using the Hartree approximation for the 
electron-electron Coulomb interaction 
\cite{Gudmundsson95:17744,Gudmundsson91:12098}. 
The spin degree of freedom is neglected. 
The circular symmetric lateral confinement
potential for the 2DEG in the quantum dot is 
\begin{equation}
      V_{\mbox{conf}}(r)=\frac{1}{2}\hbar\omega_0\left[\left(
      \frac{r}{l_0}\right)^2+a\left(\frac{r}{l_0}\right)^4\right] ,
      \label{dot_conf}
\end{equation}
where $\hbar\omega_0$ is the characteristic energy of the parabolic part
of the potential and the confinement length is 
$l_0=\sqrt{\hbar /(m^*\omega_0)}$. The dielectric constant and the effective
mass of an electron are denoted by $\kappa$ and $m^*$, respectively.
The confinement in the third direction, the $z$-direction, is considered
strong enough to justify treating the system as exactly two-dimensional.
The combined effects of the perpendicular homogeneous external magnetic field 
${\mathbf B}=B{\mathbf\hat z}$ and $V_{\mbox{conf}}$ produce the 
effective frequency of the dot 
$\tilde\omega =(\omega_c^2+4\omega_0^2)^{\frac{1}{2}}$ and the effective
length $\lambda =[\hbar /(m^*\tilde\omega)]^{\frac{1}{2}}$, replacing the
cyclotron frequency $\omega_c=eB/(m^*c)$ and the magnetic length 
$l_c=[\hbar /(m^*\omega_c)]^{\frac{1}{2}}$, respectively.
The single electron Hartree energies 
$\varepsilon_{n,M}$ and the states $|n,M)$ are
labeled by the Landau-band (LB) index $n=0,1,2,\cdots$ and the 
angular quantum number $M=-n,\cdots ,0,1,2,\cdots$,

The confinement potential for the quantum wire extended 
in the $x$-direction is of the form
\begin{equation}
      V_{\mbox{conf}}(y)=\frac{1}{2}\hbar\omega_0\left[\left(
      \frac{y}{l_0}\right)^2+a\left(\frac{y}{l_0}\right)^4\right] .
      \label{wire_conf}
\end{equation}
The effective frequency for the transverse motion of the 2DEG is
$\Omega =(\omega_c^2+\omega_0^2)^{\frac{1}{2}}$ and the effective width
is $L=[\hbar /(m^*\Omega)]^{\frac{1}{2}}$. The single electron 
Hartree energies $\varepsilon_{n,y_k}$ and the states $|n,y_k)$ are labeled
by the Landau-band index $n=0,1,2,\cdots$ and the center coordinate
$y_k=kL^2=2\pi L^2N/L_x$, 
where $N$ is an integer and $L_x$ is the length of the wire in 
the $x$-direction. The coefficient $a$ determines the deviation from 
the parabolic confinement.

The absorption of the quantum dot \cite{Gudmundsson91:12098}
and wire \cite{Gudmundsson95:17744,Brataas96:jp} is calculated 
as a linear response to the self-consistent electrostatic potential
$\phi_{\mbox{sc}}=\phi_{\mbox{ext}}+\phi_{\mbox{ind}}$, where 
$\phi_{\mbox{ind}}$ is the induced potential and $\phi_{\mbox{ext}}$
is the external potential. In case of the quantum wire 
$\phi_{\mbox{ext}}({\mathbf r},t)=y{\cal E}_{\mbox{ext}}\exp(-i\omega t)$ 
and for the quantum dot $\phi_{\mbox{ext}}({\mathbf r},t)=
r{\cal E}_{\mbox{ext}}\exp(-iN_p\varphi-i\omega t)$. The external
electrostatic field is linearly polarized transverse to the quantum wire
but in case of the quantum dot it is circularly polarized with
the choice $N_p=\pm 1$. The power absorption of the systems is calculated
from the Joule heating of the 2DEG caused by $\phi_{\mbox{sc}}$.

Kohn's theorem states that if the confinement potential is parabolic
and the external electric field is spatially homogeneous then only center
of mass motion can be excited 
\cite{Kohn61:1242,Maksym90:108,Bakshi90:7416,Pfannkuche93:6}. 
In the wire the dispersion is then
\begin{equation}
      \Omega_p = \Omega = \sqrt{\omega_c^2+\omega_0^2} ,
\end{equation}
and for the dot the dispersion has two branches 
\begin{equation}
      \omega_{\pm} = \frac{1}{2}\sqrt{\omega_c^2+4\omega_0^2}
                     \pm\frac{\omega_c}{2} ,
\label{QD_disp}
\end{equation}
with $\omega_{-}$ corresponding to the polarization $N_p=+1$ and
$\omega_{+}$ to the choice $N_p=-1$.
\section{Results of model calculation}
The calculations for the quantum dot have been performed using GaAs parameters,
$m^*=0.067m_0$ and $\kappa =12.4$. The results for 
a confinement frequency $\omega_0 =3.37\:$meV are shown in Fig.\ \ref{QD_3sub}.
The number of electrons, $N_s=60$, guarantees that states in more 
than one Landau band are occupied for $1<B<6\:$T. The dispersion of the
$\omega_-$ mode ($N_p=+1$) shows slight oscillations that correlate
with the average filling factor of the 2DEG or the location of the 
chemical potential $\mu$ with respect to the Landau bands. 
The density profiles of the 2DEG for $B=3.7\:$T and $4.9\:$T corresponding,
respectively, to a minimum and a maximum in the oscillations of 
the $\omega_-$ mode are presented in Fig.\ \ref{QD_3sub}a. 
The radial electron density for the whole range of the magnetic field 
considered here is seen in Fig.\ \ref{QD_3D_dens}. 
Clearly visible are the ``layers'' corresponding to electrons in specific
Landau bands, that are separated by sharp steps in larger quantum dots 
in a strong magnetic field, i.e.\ incompressible regions separated by 
narrow compressible regions. In the relatively small quantum dot 
considered here the steps are not quite so exact due to screening effects,
finite temperature, and the fact that the effective magnetic length $\lambda$
is not that very much smaller than the size of the quantum 
dot $R\sim 100\:$nm. For integer values of the average filling factor $\nu$
the chemical potential $\mu$ lies between the bulk parts of two Landau bands
as is seen in Fig.\ \ref{QD_E}a, otherwise $\mu$ lies in the bulk states
of a particular Landau band as Fig.\ \ref{QD_E}b shows. Comparison of
Fig.'s \ref{QD_3sub}-\ref{QD_E} shows that the oscillations of the
$\omega_-$ mode take a maximum value for an integer $\nu$ and concurrently
the steps in the density are clear. The oscillations of the $\omega_-$ mode
with $\nu$ have been explained in terms of an oscillating radius of the
electronic system \cite{Darnhofer96:xx}, an effect due to the screening 
properties of the system. We prefer to use the picture of compressible and
incompressible states to explain the oscillations \cite{Bollweg96:2774}.
For the $\omega_-$ mode these pictures are probably equivalent 
for intermediate sized 
dots, but the latter one is more convenient in case of other modes to be 
discussed here. Kohn's theorem is not programmed explicitly into the 
numerical calculation but in the density response function all 
relevant single electron-hole transitions are summed over and in the case
of a parabolic confinement only the center of mass modes predicted by the
theorem are visible. Slight deviations from the parabolic confinement
result in other modes that in the case of few electrons 
\cite{Pfannkuche94:1221} can be traced back to certain single electron
transitions or groups thereof. The dipole active collective 
modes of the 2DEG in a 
quantum dot are such that only transitions of electrons observing
$M\rightarrow M-1$ contribute to the $\omega_+$ mode, such that just
below $\mu$ a hole state with quantum number $M$ is formed but above
$\mu$ an electron state with $M-1$ is formed. An inspection of 
Fig.\ \ref{QD_E} shows that the single electron transitions involved 
are almost exclusively interband transitions satisfying $n\rightarrow n+1$.
The $\omega_-$ mode, on the other hand, has strong contributions from
intraband $M\rightarrow M+1$, hence it's much lower energy. 
This is also supported by the induced density which for the 
$\omega_-$ mode is usually concentrated close to the boundary of the
quantum dot where the concerning $M$ states have their heaviest weight.
Now the different properties of the $\omega_-$ mode with respect to an
integer filling factor or not should be clear; When $\nu$ is not close to
an integer $\mu$ is pinned to the bulk states of a particular Landau
band, where the band rises above $\mu$ its slope is much less than
at a crossing point with edge states with higher $M$. 
The single electron transitions contributing from this area thus lower
the total energy of the collective mode. The oscillations of the
$\omega_-$ mode take a maximum for $\nu\sim\mbox{integer}$ and a minimum
for half integers. The states pinned to $\mu$ are compressible and the
induced density for the $\omega_-$ mode is concentrated around the narrow
stripes of compressible states close to the edge of the dot or close
to the edge of a broader region of compressible bulk states in a 
quantum dot with not all bulk states of a particular band filled.  
The oscillations of the $\omega_-$ mode are thus a direct consequence 
of the formation of compressible and incompressible stripes.

The main contribution to the $\omega_+$ mode comes from transitions 
of occupied bulk states ($n,M$) to empty ($n+1,M-1$) states. 
The energy of these interband transitions oscillates weakly with
the filling factor, but the transitions all add up to the collective
motion that can be identified as a rigid  oscillation 
of the center of mass of the
2DEG, thus, the dispersion of the main branch of the $\omega_+$ mode shows
only much weaker oscillations than are present in the $\omega_-$ mode. 
Fig.\ \ref{QD_P1m} shows the dispersion of 
the $\omega_+$ mode for a parabolic confinement potential with the same
deviation as was used for the calculation of the $\omega_-$ mode 
in Fig.\ \ref{QD_3sub}. Around $B=2\:$T the $\omega_+$ mode shows an
anticrossing behavior just left of the line $E=2\hbar\omega_c$, the
magnetoplasmon is interacting with the first harmonic of the cyclotron
resonance, a Bernstein mode \cite{Gudmundsson95:17744}. The emerging lower
branch of the $\omega_+$ mode, the branch corresponding to the rigid
center of mass motion, becomes independent of $\nu$, but the higher
vanishing mode, that takes on the character of a higher order magnetoplasmon,
oscillates with $\nu$. The quartic deviation to the parabolic confinement
strengthens the contribution of single electron transitions of the form 
($n,M$)$\rightarrow$($n+2,M-1$) to the collective oscillations,
thus introducing the higher order magnetoplasmons that are blocked by Kohn's
theorem in a parabolically confined 2DEG. The oscillations of the side
branch of the $\omega_+$ mode (just above the main branch) 
are in antiphase to the oscillations of the
$\omega_-$ mode, explanation can be found in Fig.\ \ref{QD_E}. Without
an interaction between the electrons the Landau bands would be parallel,
and thus equidistant, interaction and screening properties of the 2DEG
change this. When $\mu$ lies in the bulk states of a particular Landau
band, i.e.\ a large compressible region is present in the density, then
each Landau band develops a flat region, but due to the increasing spatial 
extent of the wave functions of the higher bands the flat regions shrink
with growing $n$. The relative distance between the Landau bands has therefore
a maximum when the a band is pinned to $\mu$ and a large compressible
region in the electron density is present. The finite size of the 
wave functions that increases with $n$ also explains why the ``layers''
in the density of a relatively small quantum dot get thinner and smaller
with increasing $n$. A large quantum dot does not exhibit this effect.

The observation of higher order dipole active magnetoplasmons opens the
question what effects the deviation from the parabolic confinement has
on the quadruple active magnetoplasmon
\cite{Gudmundsson91:12098,Glattli85:1710,Merkt91:7320},
$N_p=\pm 2$, even though it has not been 
measured directly in quantum dots. This collective mode has contributions
of single electron transitions with $\Delta M=\pm 2$ and is seen in 
Fig.\ \ref{QD_modes} together with the dipole active modes discussed
above. The quadrupole $\omega_+$ mode shows a very clear and simple
$2\omega_c$ anticrossing, much simpler than in the dipole case.
The quadrupole active $\omega_-$ mode shows stronger filling factor
oscillations than the dipole counterpart and the oscillations are in phase
since they have the same origin. In the lower split-off branch of the
quadrupole active $\omega_+$ mode weak oscillations with $\nu$
are found in phase with the dipole counterpart.

In light of the results above for quantum dots it would not be unexpected
to find some filling factor dependent effects in quantum wires, even though
the magnetoplasmon in a parabolically confined system
has only one branch corresponding to the $\omega_+$ mode of a quantum dot.
The electronic density of a quantum wire develops clearly separate 
compressible and incompressible regions with increasing wire width, and
the steps are also clear in the Hartree energy spectra.  
In the case of a dipole excitation the single electron transitions 
fulfill the conservation of the center coordinate, $\Delta y_k=0$.
The calculation for the wire has been performed with the following 
parameters, $\hbar\omega_0=3.94\:$meV, $a=0.03$, and $\kappa =12.53$, other
parameters are the same as for the quantum dot. The results for the
dispersion are seen in Fig.\ \ref{QW_modes}. Again we have the $2\omega_c$
anticrossing, now just right of the $E=2\hbar\omega_c$ line. For $B>3\:$T
the lower branch gains oscillator strength and corresponds to the rigid
center of mass motion of the whole 2DEG. The upper branch is a higher order
magnetoplasmon as the induced density confirms. The dispersion has a very
similar overall appearance as the dispersion for the $\omega_+$ mode 
of the quantum dot seen in Fig.\ \ref{QD_P1m}, and the Hartree energy
spectra show similar properties as the spectra for quantum dots, thus
leading to the same explanation of the origin of the oscillations
with $\nu$.
\section{Summary}
We have been able to explain filling factor dependent oscillations
occurring in the magnetoplasmon dispersion for quantum dots and wires
in terms of the formation of compressible and incompressible regions
in the electronic density of the systems. The oscillations have been 
found in the $\omega_-$ mode of quantum dots \cite{Bollweg96:2774} and
some indications of them have been found in wires and in the
$\omega_+$ mode for dots. Their appearance in wires and in the $\omega_+$ 
mode for quantum dots seems to depend on the concerning system to be
of intermediate size, i.e.\ the effective magnetic length needs to
be not much less than one order of magnitude smaller than the system.
The far-infrared measurements are therefore yet another method to 
observe the internal structure of not parabolically confined quantum
dots and wires, the stripes formed by the compressible and incompressible 
states.

\acknowledgements{This research was supported in part by the Icelandic
Natural Science Foundation, the University of Iceland Research Fund,
and a NorFA Network Grant.}
\bibliographystyle{prsty}

%
%
%
\begin{figure}
\caption{(a) Results of the  Hartree approximation for the 
         density profile vs.\ the spatial coordinate $x$ in a quantum dot 
         in units of $10^{11}\:$cm$^{-2}$. The flat regions for 
         $B=4.9T$ indicate the incompressible regimes.
         (b) Resonance frequency of the $\omega_-$ mode in RPA. 
         (c) Ratio of the RPA and the classical 
         result (\protect{\ref{QD_disp}}),
         $\omega_-^{RPA}/ \omega_-$, vs $B^{-1}$. $T=1\:$K, 
         $m^*=0.067m_0$,
         $\kappa =12.4$, $\omega_0 =3.37\:$meV, and $a=0.0674$. }
\label{QD_3sub}
\end{figure}
\begin{figure}
\caption{The radial electron density in units of $10^{11}\:$cm$^{-2}$ 
         for a quantum dot in the
         Hartree approximation as a function of the magnetic field $B$
         and the radius $r$. Same parameters as in 
         \protect{Fig.\ \ref{QD_3sub}}.}   
\label{QD_3D_dens}
\end{figure}
\begin{figure}
\caption{The Hartree single-electron energy spectra vs the angular momentum
         quantum number $M$ for (a) $B=4.9\:$T, and (b) $B=5.7\:$T. 
         Same parameters as in \protect{Fig.\ \ref{QD_3sub}}.}
\label{QD_E}
\end{figure}
\begin{figure}
\caption{The power absorption $P(\hbar\omega )$ for the $\omega_+$
         mode ($N_p=-1$) of quantum dot. 
         The calculation was performed for $B\sim 1-6\:$T.
         A constant Lorentzian linewidth
         $\hbar\eta =0.01\omega_0\:$meV is used.
         Other parameters
         are as in \protect{Fig.\ \ref{QD_3sub}}.}
\label{QD_P1m}
\end{figure}
\begin{figure}
\caption{The dispersion of the quadrupole modes 
         in a quantum dot labelled with 
         $N_p=-2$ and $N_p=+2$, and the dipole modes labelled with
         $N_p=-1$ ($\omega_+$)and $N_p=+1$ ($\omega_-$).
         The straight lines represent the energies
         $E=\hbar\omega_c$, and $E=2\hbar\omega_c$.
         The calculation was performed for $B\sim 1-6\:$T.
         Other parameters are as
         in \protect{Fig.\ \ref{QD_3sub}}.}
\label{QD_modes}
\end{figure}
\begin{figure}
\caption{The dispersion of dipole modes in a quantum wire
         with $a=0.03$.  The one dimensional electron density
         $n_{1D}=1.25\times 10^{6}\:$cm$^{-1}$, $T=1\:$K, 
         $m^*=0.067m_0$, $\kappa =12.53$, and $\omega_0 =3.94\:$meV. }
\label{QW_modes}
\end{figure}
\end{document}